# Single-Item Fashion Recommender: Towards Cross-Domain Recommendations

Seyed Omid Mohammadi
*University of Tehran*
*College of Engineering, School of Electrical and Computer Engineering*
Tehran, Iran
S.OmidMohammadi@alumni.ut.ac.ir

Hossein Bodaghi
*University of Tehran*
*College of Engineering, School of Electrical and Computer Engineering*
Tehran, Iran
Hossein.Bodaghi@ut.ac.ir

Ahmad Kalhor
*University of Tehran*
*College of Engineering, School of Electrical and Computer Engineering*
Tehran, Iran
AKalhor@ut.ac.ir

*Abstract*— Nowadays, recommender systems and search engines play an integral role in fashion e-commerce. Still, many challenges lie ahead, and this study tries to tackle some. This article first suggests a content-based fashion recommender system that uses a parallel neural network to take a single fashion item shop image as input and make in-shop recommendations by listing similar items available in the store. Next, the same structure is enhanced to personalize the results based on user preferences. This work then introduces a background augmentation technique that makes the system more robust to out-of-domain queries, enabling it to make street-to-shop recommendations using only a training set of catalog shop images. Moreover, the last contribution of this paper is a new evaluation metric for recommendation tasks called objective-guided human score. This method is an entirely customizable framework that produces interpretable, comparable scores from subjective evaluations of human scorers.

## I. INTRODUCTION

The ever-growing fashion e-commerce has led to a massive increase in the number of fashion items posted online. However, the bigger the market, the harder it will be for the customers to find items suited to their needs. Thus, online shops are beginning to implement visual search engines and recommender systems.

This article takes three steps towards achieving a single-item fashion recommender system that can handle both in-shop and street-to-shop recommendations. The first step aims to design and develop a content-based recommender system using fashion items' "Shop" images. The proposed system uses deep convolutional neural networks to extract visual features of images and forms recommendations based on a similarity check. Then, we utilize this content-based structure to create a two-stage personalized recommender system and show that the proposed method improves previous works' results. The next stage of this research is dedicated to finding a solution that tackles the problem of cross-domain recommendation using a background augmentation technique.

While evaluation methods for retrieval tasks are numerous and well-developed, including precision/recall @k and accuracy, it is challenging to define a precise objective evaluation metric for recommendation tasks. One can only use the same retrieval evaluation metrics for these tasks if similar items are labeled and ground truth is available, which is not most of the time. Thus, we also propose an evaluation framework to obtain quantitative information from subjective evaluations and be able to compare multiple recommendation systems with a specific goal in mind.

The main contributions of this article are as follows:

- It develops an in-shop content-based fashion recommender and shows its power in returning similar fashion items from single catalog image queries.

- It utilizes previous works to turn this system into a personalized fashion recommender that considers user preferences. An increase in performance is also illustrated.

- It presents a new background augmentation technique to narrow the gap between Shop and Street fashion image domains and develop a system much more robust to out-of-domain queries.

- Finally, it suggests a method of evaluating recommender systems with specific goals and discusses the evaluation results of several systems.

## II. RELATED WORKS

Reference [1] used a convolutional neural network, VGG, to be specific and cosine similarity for content-based recommendations. However, the network was trained using less than 50 thousand images, and it could only classify the items into seven single-label categories. On the other hand, [2] experimented with AlexNet and BN-Inception mixed with KNN as similarity measurement. However, the number of classes was still limited to nine categories and five texture types classified using two separate networks. Like many other studies, this work did not compare the results of the recommendation task and only used classification accuracy as an implicit objective metric.

For personalization purposes, [3] proposed a structure using DenseNet, pre-trained on ImageNet, to capture user preferences and recommend a set of outfits for each user. Moreover, inspired by the structure of VisNet [4], [5] used ResNet101 paralleled with a shallow net to extract features and generated personalized recommendations using a second dense neural network. This study used ResNet101 and showed that the results were superior compared to that of VGG16. Furthermore, unlike previous works, the proposed network used multi-label classification, and the number of classes was considerably high, which was closer to real-life conditions.

A well-known challenge in fashion recommendation is the gap between professional catalog images used in online shops (called Shop images) and user-created images (called Street or Wild images). Reference [6] was one of the firsts to address this issue. Soon others gained interest as [7] used visual phrases and [8] used articulated pose estimation and image retrieval techniques. Deep learning-based approaches such as [9] and [10] led to far superior results. Zalando researchers proposed a model called Street2Fashion to segment the background of images in 2019, which highly improved their results [11]; unfortunately, they did not release their images. Furthermore, recent studies suggest that it is possible to further improve the results by using fashion landmarks [12], [13].

While most of the studies mentioned above use two sets of data (Shop and Street) to train their networks, these cross-domain fashion datasets are rarely available. It can be challenging to find a large-scale, publicly-available fashion dataset that meets the needs of training such systems as most of the existing datasets are either too small (in terms of size or number of tags) or not accessible to the public. Thus, we aim to put the idea, whether it is possible to achieve acceptable results using only one dataset of Shop images, to test. We propose a background augmentation technique to use a Shop dataset and simulate the conditions of Wild or Street fashion datasets.

## III. SINGLE-ITEM CONTENT-BASED RECOMMENDER

The structure of the proposed content-based recommender is illustrated in Fig.1. The feature extractor part follows the work of [5], which is a multi-label classifier, with some modifications. We change the network from ResNet101 to ResNet50, which improves the system's efficiency without harming the final results. The loss function for training the network (1) is the same weighted cross-entropy used in [5]. It uses the frequency of each label $i$ ($f_i$) in the set of all classes ($C$) to determine the weight of that label ($\omega_i$) and uses these weights to calculate the loss between the target ($t_i$) and the output ($o_i$) of the system. We add a minor trick according to (3) in which $F1$ (F-score) and IOU are not quantized values. This change helps further balance the results in terms of precision and recall and also gradually reduces the loss through the process as the system becomes stable and minute weight changes are required.

$$loss(o,t) = -\frac{1}{n}\sum_i \omega_i(t_i.\log(o_i) + (1-t_i).\log(1-o_i)) \quad (1)$$

where:

$$\omega_i = \left(\max_{j \in C} f_j + 0.01\right) - f_i \quad (2)$$

and

$$Loss = loss(o,t) \times (1 - F1) \times (1 - IOU) \quad (3)$$

The final feature extractor is depicted in Fig.2. It takes a single fashion item catalog image as input and predicts labels for that image from a pool of 1102 labels. The features are extracted from the layer before the last (classification) layer. Now that visual features are obtained, a similarity check is needed to form the recommendation list. Cosine similarity is used here as it is fast and leads to great results.

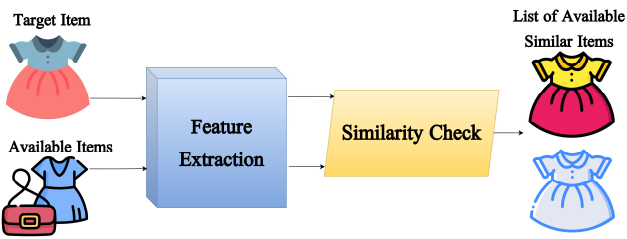

Fig.1. Structure of the content-based fashion recommender

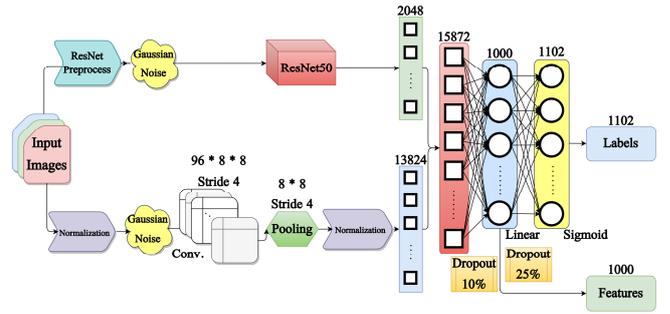

Fig.2. The feature extractor network (multi-label classification task)

The next phase is to personalize the recommendations. The same feature extractor network can be used for this purpose. For the personalization network, we use the structure proposed by [3]. The system can take a mix of previously purchased items' image features (called Cart or shopping cart items) as input and recommend a list of similar items tailored to a specific user's taste. This mix of items can be an average of all items' features weighted by user rating as in (4). This way, there will not be a limit to the number of items before a prediction can be made, as this method supports carts with any number of items.

Furthermore, this way, we explicitly use user preferences in the form of user ratings. The main downside to this is that carts should be uniform and from the same fashion categories. One way is to break a user's main cart into multiple smaller uniform carts and use the network to find recommendations for each group separately. For example, one can use K-Splits [14] to automatically identify different item categories formerly purchased by a user (namely shoes, shirts, jeans) and cluster similar items together. There are many ways to improve this system, but personalization is not the main focus of this article.

$$Cart = \frac{\sum_{i \in Bought}(Rating_i \times Features_i)}{\sum_{i \in Bought} Rating_i} \quad (4)$$

### A. Towards Street-To-Shop Recommendations

Previous works which used deep domain adaptation methods utilized images from two different domains. However, this research aims to train a network robust to street images by only using one fashion dataset made of Shop catalog images with neutral backgrounds and standard poses. Thus, this study proposes a new augmentation technique to create street-like images that force the network to focus on fashion items present in the image and not the background.

The background augmentation technique puts fashion items in front of various backgrounds with different levels of complexity in the training phase of the network. For this purpose, we find the boundaries of the fashion item or model shown in the image by using contours. This task will generally be straightforward and error-free as shop images contain neutral and one-color backgrounds. After this, the main images are masked and cropped from the background. Next, we paste these masks onto random, more complex backgrounds. This method has several benefits. Firstly, the system sees the same fashion outfit in front of varying real-world backgrounds in each run, which helps focus attention only on the outfits and not the whole image. Secondly, we have complete control over the whole process. Lastly, a suitable dataset of real-world scenes and images can be gathered as backgrounds with minimum effort. Fig.3 shows images with standard augmentation techniques, including flip, rotation, shift, and shear, alongside images with background augmentation. Using this method can close the gap between fashion shop images and street images.

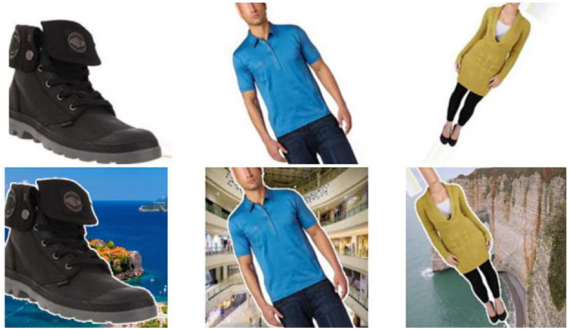

Fig.3. Images with regular augmentation (top row) and with background augmentation (bottom row)

## IV. Experiments

The experiments are conducted using Python 3.7.4 and Tensorflow and Keras libraries. This work needs a large-scale dataset of "Shop" fashion images with user history and item descriptions. Thus, it chooses the Amazon Fashion dataset [15] as it perfectly suits its needs.

### A. Evaluation Methods

Metrics like accuracy are not applicable in evaluating multi-label multi-class classifications with huge numbers of classes. Thus, we use Intersection Over Union (IOU) to evaluate the label prediction task. Although IOU is more prevalent in evaluating object detectors, it provides excellent insight for multi-label classification tasks as well because it can measure the overlap of predicted and actual labels if formulated as (5).

$$IOU = \frac{TP}{TP + FP + FN} \quad (5)$$

However, evaluating the recommendation results is not that straightforward as subjectivity plays an essential role in it, and there is no ground truth. Logically, better visual features lead to better recommendations. Thus, the IOU or feature extractor's classification accuracy can be used as implicit objective metrics for this purpose, but these are not one-to-one comparable. This article tackles this issue by proposing an objective-guided human score which turns a subjective evaluation of the results into comparable numerical values based on specific goals of the recommendation system.

Not all recommender systems are the same. In fact, the goal behind using such systems might be completely different from company to company. If the system is used as an image search engine, the results are expected to be of the same category with similar shapes and colors. On the other hand, recommendation results for advertising need to be novel items with a wide variety. Hence, an objective-guided metric is proposed to evaluate different systems based on the goals they are meant to satisfy.

First, for an objective-guided human score, some fashion images are needed as queries. These images should be carefully chosen to indicate the goal of the system. For example, the ratio of Shop/Street images should be set based on the predicted ratio of Street image queries the systems need to handle in the future. It is good to mix a set of easy and more complex images to make sure the system works well under different circumstances. The number of images is also optional; more images probably increase the accuracy of this evaluation, provided that they do not bore or tire the human scorers. In short, the number of chosen images, their domain, and their complexity are all controllable based on the system's goals.

Next, multiple criteria are defined to score the functionality of each system based on them. This step, again, is entirely flexible based on the needs of the evaluators. This article uses seven criteria: category, subtype, fabric/texture, color, variety, details, and shape difference. The order of showing the results, domain sensitivity, price range, and many more criteria can be added to this list optionally.

- **Category:** Defines the main category of an image, such as top, bottom, footwear, and jewelry.
- **Subtype:** Defines subtypes of the same category, such as boots, high heels, college, and slippers.
- **Fabric/Texture:** Shows the main fabric or garment's texture, such as denim, leather, smooth, and shiny.
- **Color:** Defines the dominant color of the item, such as red, green, blue, yellow.
- **Variety:** The number of novel items (different category, subtype, or color). Almost on the opposite side of the other criteria, because the higher the variety score is, the lower the other scores.
- **Details:** The number of results that follow fine details, such as necklines, zipper, pockets, and design.
- **Shape Difference:** The number of items that do not follow the outline of the query item, such as images with different angles, different perspectives, rotations, and flips.

Finally, we input the queries into all different recommender systems, save top-10 results for each, and create an evaluation sheet. Every human scorer now scores each system separately based on individual criteria mentioned before. For each criterion, the scorer assigns a hit@10 score from 0 to 10. These scores can easily be turned into percentages later. All criteria scores are then processed using a weighted average, and the weight of each criterion is set based on its importance regarding the goal of the system. Additionally, all the scores for each system are averaged as well, and one final objective-guided human score for each system is obtained, which will be comparable to other scores provided that they have the same goals.

This method is called objective-guided as almost all of its parameters (queries, criteria, and weights) are flexible and can be set based on the needs of the evaluator. Nevertheless, once the parameters are set for a specific goal, the results of this evaluation will be directly comparable. Not only that, but evaluation results of each criterion are also comparable, which helps determine the effects of different methods on each criterion separately and increases interpretability.

### B. In-Shop Content-Based Recommendations

For this purpose, the network shown in Fig.2 is used. The dataset is separated into 80% train, 10% test, and 10% validation. All images are resized to $224 \times 224$ with the aspect ratio unchanged. ResNet50 is loaded with ImageNet pre-trained weights, and the whole network is fine-tuned using the training data at hand. Dropout, regularizations, early stopping, and regular augmentation techniques are used to improve the results. The final test results of the trained system are 44.4% IOU, 73.1% precision, and 48.4% recall for the multi-label classification task.

Next, two random images are chosen as queries, and after feeding them through the network, cosine similarity is used to form Top-8 recommendations. As shown in Fig.4, the results are fantastically similar in all aspects, including color, category, and design.

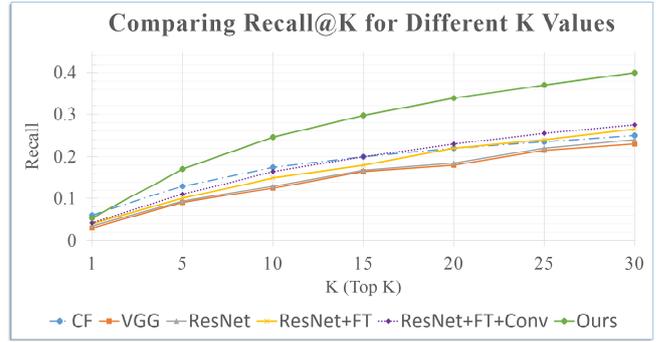

Fig.5. Recall@K comparison of different methods

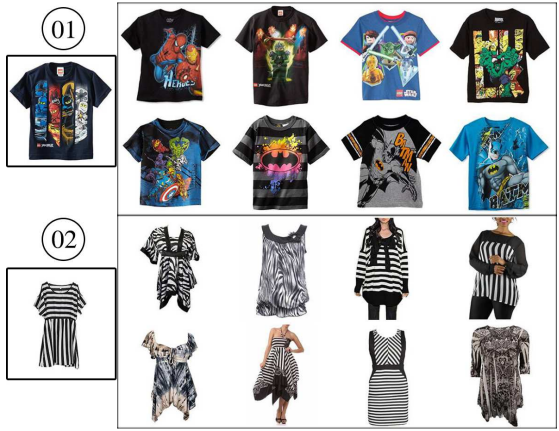

Fig.4. Top-8 recommendation results for two random image queries

### C. Personalized Recommendations

This section trains a second network on top of the feature extractor trained in the previous section to personalize the results. A total of 44,103 users are chosen, each with 5-7 reviews to balance the data. The total number of reviews is 269,104, so each user has approximately 6.1 reviews (bought items). Carts are calculated based on (4) using these users, their bought items, and ratings.

After training the network, it reaches 78.4% test accuracy. In Fig.5, the proposed method (fine-tuned ResNet50 paralleled with a shallow net and a personalizer network on top) is compared to collaborative filtering, VGG, ResNet101, fine-tuned ResNet101, and fine-tuned ResNet101 paralleled with a shallow net. The Recall@K result for the proposed method is almost always better than previous methods for all K values, except for Top-1 recommendation results.

### D. Street-to-Shop Recommendations

The network is robust to some degree of change, but this robustness is not enough to enable the system to handle out-of-domain images efficiently. Thus, this section re-trains the network, this time using the background augmentation technique introduced in Sec. III.A.

A dataset of several thousand backgrounds is collected, and the system randomly chooses a background for each image in each epoch. After completing the training process, the system can handle both in-shop catalog images and more complex street-like queries. Fig.6 shows the results for out-of-domain examples from Instagram online shops.

Five networks are compared in this section. These versions all utilize the background augmentation and the same training techniques and are as follows. **V1** is only a ResNet50 network, **V2** is a ResNet50 paralleled with a shallow network, **V3** is similar to V2 but with heavier augmentations to check to what extent traditional augmentation techniques can improve street-to-shop recommendations, **V4** is only an EfficientNet [16] with $300 \times 300$ input, and **V5** is the same EfficientNet paralleled with a shallow network.

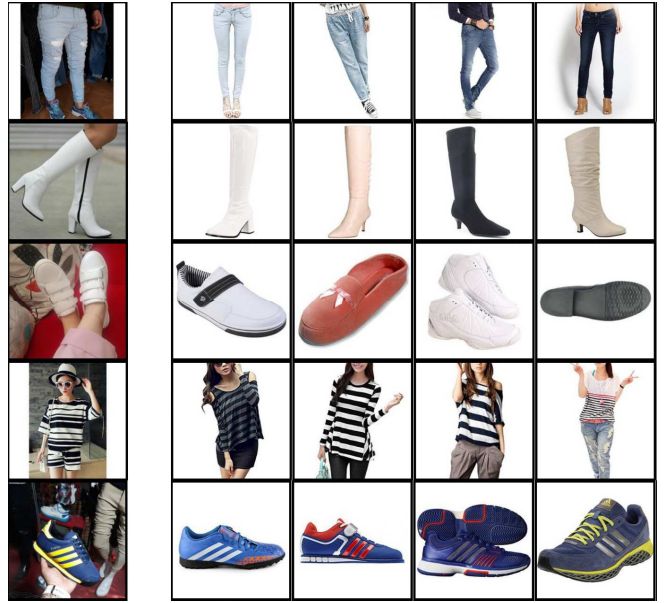

Fig.6. Top-4 street-to-shop recommendation results for five random out-of-domain queries

Seven criteria, as explained in Sec. A, are chosen to compare these systems. Fifty query images with mixed complexities and a Shop/Street ratio of 2/3 are prepared and fed to each system, resulting in 50 evaluation sheets. Next, human evaluators rate (hit@10) the results of each system for each query based on each criterion separately. The scores are then processed using a weighted average based on the importance of each criterion for the specific goal of the network. The scores for each criterion and the final objective-guided human score (OHS) for each system are provided in TABLE I. in percentages. The best and worst scores in each row are in green and red colors, respectively.

TABLE I. reveals fantastic information about the systems. First of all, it can be seen that although the IOU of the classification task of the feature extractor can be used as an implicit evaluation of the recommendation system, it is not directly comparable. IOU is also not interpretable either. On the other hand, OHS explicitly shows the effect of each parameter change on every single criterion.

Adding a shallow branch to deep networks marginally decreases the category score but improves color and variety scores. This finding agrees with [5] as they pointed out that a shallow net helps recover the color, which is usually lost in deeper network structures. Additionally, this technique leads to better results in terms of final OHS. The power of EfficientNet is also shown as both V4 and V5 output far more superior results. It seems that variety and shape difference are the only criteria in which the EfficientNets show weakness. Thus, it is preferable to use these networks in image search engines in which these two criteria are of less importance.

## V. Conclusion and Future Work

Fashion e-commerce is growing at an unbelievably fast pace. However, as the number of fashion items in online shops grows exponentially, the difficulty of finding specific items in this vast market increases.

This article provided a single-item content-based fashion recommender for in-shop search and recommendation by training a convolutional neural network consisting of double branches, a deep ResNet50 network, and a shallow net. Then, it used this network as a feature extractor in a two-stage structure to form personalized recommendations. Next, a background augmentation method was proposed to enable the system to make cross-domain suggestions. Finally, this work introduced objective-guided human score, an evaluation metric for recommendation tasks, a customizable framework that turns subjective evaluations into interpretable and comparable percentage values. The comparison results showed that EfficientNets could be superior to ResNets for the task at hand.

One future research direction is to utilize clustering methods like k-means and k-splits to guide the personalization system through the fashion feature space. Another promising direction is using advanced natural language processing methods and relation extraction techniques to obtain noise-free labels. Furthermore, a comparison of the method with a twin or triplet network would also be scientifically interesting.

TABLE I.     Objective-guided Human Score (OHS) comparison of five networks

| Version No. | V1 | V2 | V3 | V4 | V5 |
|---|---|---|---|---|---|
| Test IOU | 44.60% | 43.90% | 44.50% | 45.20% | 44.10% |
| Category | 96.90% | 93.70% | 95.40% | 99.80% | 98.30% |
| Subtype | 71.50% | 70.40% | 71.70% | 77.30% | 86.30% |
| Texture | 68.50% | 77.70% | 69.20% | 86.50% | 80.00% |
| Color | 67.20% | 68.30% | 64.80% | 70.20% | 73.50% |
| Variety | 24.80% | 32.70% | 33.30% | 21.70% | 27.30% |
| Details | 36.30% | 43.50% | 33.30% | 43.40% | 37.90% |
| Shape Difference | 32.70% | 31.70% | 20.10% | 20.20% | 21.00% |
| OHS | 73.60% | 74.00% | 73.20% | 78.10% | 80.00% |